\documentclass[pra,twocolumn, preprintnumbers,superscriptaddress]{revtex4} 

\usepackage{latexsym,epsfig,amssymb,amsfonts,amsmath,graphicx,bbm}

\usepackage{color}
\definecolor{dred}{rgb}{.8,0.2,.2}
\definecolor{dyellow}{rgb}{.7,.7,.0}
\definecolor{ddred}{rgb}{.4,.0,.0}
\definecolor{dblue}{rgb}{.2,.2,.8}
 
\newcommand{\half}{\mbox{$\textstyle \frac{1}{2}$}}
\newcommand{\ket}[1]{ |  #1 \rangle}

\newcommand{\bra}[1]{ \langle #1  |}
\newcommand{\braket}[2]{\left\langle\, #1\,|\,#2\,\right\rangle}

\newcommand{\eqr}[1]{Eq.\ (\ref{#1})}
\newcommand{\fir}[1]{Fig.\ \ref{#1}}

\newcommand{\an}[1]{\hat{#1}}
\newcommand{\cre}[1]{\hat{#1}^\dag}

\begin{document}

\title{Solving search problems by strongly simulating quantum circuits}

\author{T. H. Johnson}
\email{t.johnson1@physics.ox.ac.uk} 
\affiliation{Clarendon Laboratory, University of Oxford, Parks Road, Oxford OX1 3PU, United Kingdom}
\author{J. D. Biamonte}
\affiliation{Centre for Quantum Technologies, National University of Singapore, 3 Science Drive 2, 117543, Singapore}
\affiliation{Institute for Scientific Interchange, Via Alassio 11/c, 10126 Torino, Italy}
\author{S. R. Clark}
\affiliation{Centre for Quantum Technologies, National University of Singapore, 3 Science Drive 2, 117543, Singapore}
\affiliation{Clarendon Laboratory, University of Oxford, Parks Road, Oxford OX1 3PU, United Kingdom}
\affiliation{Keble College, University of Oxford, Parks Road, Oxford OX1 3PG, United Kingdom}
\author{D. Jaksch}
\affiliation{Clarendon Laboratory, University of Oxford, Parks Road, Oxford OX1 3PU, United Kingdom}
\affiliation{Centre for Quantum Technologies, National University of Singapore, 3 Science Drive 2, 117543, Singapore}
\affiliation{Keble College, University of Oxford, Parks Road, Oxford OX1 3PG, United Kingdom}

\date{\today}

\begin{abstract}
Simulating quantum circuits using classical computers lets us analyse the inner workings of quantum algorithms. 
The most complete type of simulation, strong simulation, is believed to be generally inefficient.
Nevertheless, several efficient strong simulation techniques are known for restricted families of quantum circuits and we develop an additional technique in this article.
Further, we show that strong simulation algorithms perform another fundamental task: solving search problems.
Efficient strong simulation techniques allow solutions to a class of search problems to be counted and found efficiently.
This enhances the utility of strong simulation methods, known or yet to be discovered, and extends the class of search problems known to be efficiently simulable.
Relating strong simulation to search problems also bounds the computational power of efficiently strongly simulable circuits; if they could solve all problems in $\mathrm{P}$ this would imply the collapse of the complexity hierarchy $\mathrm{P} \subseteq \mathrm{NP} \subseteq \# \mathrm{P}$.
\end{abstract}

\maketitle 

Ever since Shor famously showed that a quantum computer can factorise numbers super-polynomially quicker than the fastest known classical algorithm~\cite{Shor1994}, researchers have sought methods to efficiently simulate quantum circuits using classical computers.
This is partly to narrow down where the power of quantum computers may arise and partly to understand the operation of quantum algorithms. 
Simulation is typically divided into two types, weak and strong~\cite{Jozsa2003,VandenNest2010}. Weak simulations sample an outcome of a measurement on a quantum circuit with the correct probability distribution. Strong simulations, which we focus on, go further, precisely calculating probabilities in this distribution. 

The strong simulation of an arbitrary quantum circuit is widely believed to be inefficient, even using a quantum computer; in the language of computational complexity, it is \#P-hard~\cite{VandenNest2010}. However several efficient methods are known for strong simulation using a classical computer, provided the process to be simulated is sufficiently restricted~\cite{Gottesman1999,Knill2001,Valiant2001,Terhal2002,Jozsa2003,Vidal2003,Aaronson2004,Bravyi2005,DiVincenzo2005,Anders2006,Jozsa2006,Shi2006,Yoran2006,Yoran2007,Browne2007,Bravyi2007,VandenNest2007,Jozsa2008,Clark2008,Markov2008,Bravyi2009,VandenNest2009}. For example, Gottesman showed that stabiliser circuits acting on computational basis states admit efficient strong simulation if measurements are restricted to that basis~\cite{Gottesman1999,Aaronson2004,Anders2006,Clark2008}. Valiant demonstrated a similar result for circuits comprising matchgates~\cite{Valiant2001}. This was later related to free Fermion systems and generalised to Gaussian circuits~\cite{Knill2001,Terhal2002,DiVincenzo2005,Bravyi2005,Jozsa2008}. Further, utilising techniques for contracting tensor networks, Markov and Shi developed a method for efficiently simulating circuits composed of arbitrary gates arranged in a tree-like geometry~\cite{Markov2008}.

In this article we exploit that every strong simulation of a quantum circuit is represented by the contraction of a tensor network. We discuss how individual gates are represented by tensors and the pattern of connections between the tensors follows directly from the circuit. 
Contracting the resulting tensor network returns the strong simulation result.
An advantage of the tensor network representation is that it acts as a unifying language for strong simulation algorithms, and we use it here to demonstrate that concatenations of tree-like circuits followed by Gaussian or stabiliser circuits are efficiently strongly simulable. Another advantage of using the language of tensor network contraction is that it applies beyond quantum circuits, to classical Boolean circuits~\cite{Biamonte2010} and their generalisation to stochastic circuits~\cite{Johnson2010}.

Strong simulation algorithms efficiently perform restricted forms of what is thought to be a generally inefficient task. It is then worth investigating whether they might efficiently solve subsets of other generally difficult problems. In this regard, Aaronson and Gottesman showed that solutions to problems in the class $\oplus \mathrm{L}$ (parity-$\mathrm{L}$)~\cite{Damm1990} could be embedded in the probabilities of stabiliser-circuit measurement outcomes and thus efficiently solved using methods to simulate such circuits~\cite{Aaronson2004}. Moreover, following Valiant, it has been shown that matchgate circuit simulators, and the related holographic algorithms, provide an efficient means to solve problems for which none were previously known~\cite{Cai2007}. 

This article is similarly motivated; we show that every efficient strong simulation technique provides a method for efficiently solving a class of search problems. 
We do this by constructing, for any given search problem, circuits that check solutions to the problem and whose strong simulation counts and reveals solutions to the problem. A search problem is then efficiently solvable if these strong simulations can be performed efficiently.

The importance of search problems stems from the wide variety of tasks that can be phrased in terms of them~\cite{Goldreich2010}. Further, by connecting strong simulation to search problems we highlight the relationship between strong simulation and well-known complexity classes: the problems decided by efficiently strongly simulable circuits must form a proper subset of $\mathrm{P}$, otherwise $\mathrm{P} = \mathrm{NP} = \# \mathrm{P}$.

\section*{Results}
{\bf Search problems.}
A search problem is a collection of instances, each represented by a bitstring $x$ and associated with a set of solutions $S(x)$~\cite{Goldreich2010}. Each solution $w \in S(x)$ is also represented by a bitstring. 
A search problem is efficiently solvable if, for every instance $x$, it is possible to find a solution $w \in S(x)$ or determine none exists in a time $\le p(|x|)$. Here $p$ is some polynomial and $|x|$ is the length of bitstring $x$. This is consistent with the general notion of {\em efficient}, meaning that a task can be completed in a time upper-bounded by a polynomial in the size of its description. In what follows, we assume that solutions can be efficiently checked using a classical computer and efficiently described.

{\bf Searching by counting.} 
Our approach relies on the reduction of a search problem to a counting problem~\cite{Goldreich2010}.
Each instance of the counting problem, described by $(x,n,w')$, is to calculate $\#(x,n,w')$ the number of solutions to $x$ that are of length $n$ and end in some bit pattern $w'$.

The reduction, depicted in \fir{fig:fig1}, is performed as follows. 
For a given $x$ with $\mathtt{Null}$ the zero-length bitstring, calculate $\#(x,n,\mathtt{Null})$ for $n=1$. Repeat this for $n=2,3,\dots$\ until either a non-zero value of $\#(x,n,\mathtt{Null})$ is obtained or $n$ reaches some maximum value $n_{\mathrm{max}}$. If $n_{\mathrm{max}}$ is reached without obtaining a non-zero value, then no solution exists. If instead a non-zero value is found for some $n = |w|_{\mathrm{min}} \le n_{\mathrm{max}}$, then a length-$|w|_{\mathrm{min}}$ solution exists. To find one such solution, complete the following binary search. Evaluate $\#(x,|w|_{\mathrm{min}},0)$. If this is non-zero, calculate $\#(x,|w|_{\mathrm{min}},00)$. Otherwise, calculate $\#(x,|w|_{\mathrm{min}},01)$. Continuing in this way, a length-$|w|_{\mathrm{min}}$ solution to $x$ is inferred after a total of $2|w|_{\mathrm{min}}$ counts.

Since solutions are efficiently described, we may choose $n_{\mathrm{max}} \le p(|x|)$ and this results in two properties: first, $x$ is solved by calculating $\#(x,n,w')$ for $\le p(|x|)$ values of $n$ and $w'$; second, for these $n$ and $w'$ we have $|(x,n,w')| \le p(|x|)$. 
It follows that if the counting problem is efficiently solvable then so is the search problem. 

{\bf Counting by simulating.}
To highlight the computational difficulty ($\#$P-hardness) of strong simulation, researchers have constructed quantum processes with outcome probabilities encoding the solutions to hard counting problem instances~\cite{VandenNest2010}.
We encompass such approaches by constructing a quantum process whose outcome probabilities embed the solution to an arbitrary counting instance $(x,n,w')$. Our motivation, unlike in~\cite{VandenNest2010}, is not to suggest the general inefficiency of strong simulation but to take specific cases in which it is efficient and transfer this efficiency to solving the embedded counting instance.

We now outline the quantum process, leaving the details of its construction to the methods section.
The quantum process consists of an input state, a quantum circuit and a measurement. 
Each possible solution $w$ of length $n$ is represented by an $N$-qubit product input state
\begin{equation}
\ket{w} = \ket{0}^{\otimes (N-n)} \ket{ w_{n} } \cdots \ket{ w_1 } , \label{eq:w}
\end{equation}
where $\{ \ket{0}, \ket{1} \}$ is the computational basis. Input states are evolved according to a quantum circuit $C_{x,n}$ comprising $M$ gates, with each gate $g_k$ acting on a bounded number of qubits. The unitary operator representing this circuit is
\begin{equation}
\hat{C}_{x,n} = \prod_{k=1}^M \hat{g}_k. \label{eq:circuit}
\end{equation}
Finally, a measurement is performed, defined by some projectors $\{ \hat{\Pi}, \mathbbm{1} - \hat{\Pi} \}$ with outcomes $\{ \mathrm{ {\tt yes}}, \mathrm{ {\tt no}} \}$. Here we choose $\an{\Pi}$ to project onto a computational basis state of the last qubit
\begin{equation}
\hat{\Pi} = \mathbbm{1}_2 \otimes \cdots \otimes \mathbbm{1}_2 \otimes \ket{1}\bra{1} . \label{eq:projector}
\end{equation}

The circuit $C_{x,n}$ is devised such that if $w \in S(x)$ then $\hat{C}_{x,n} \ket{w}$ is in the range of $\hat{\Pi}$, otherwise it is in the range of $\mathbbm{1} - \hat{\Pi}$. We call a circuit with this property a {\em solution-checking circuit}: if $\ket{w}$ is inputted into $C_{x,n}$ then measuring the output returns {\tt yes} with probability 
equal to unity if $w \in S(x)$, and otherwise zero. Then, by linearity, inputting a superposition
\begin{align}
\ket{W(n,w')} &= \mathcal{N} \sum_{w \in W(n,w')} \ket{w}  \nonumber \\
&= \ket{0}^{\otimes (N-n)} \ket{ + }^{\otimes (n-n')} \ket{ w'_{n'} } \cdots \ket{ w'_1 } , \label{eq:superposition}
\end{align}
into $C_{x,n}$ returns {\tt yes} with probability $\mathcal{P} = \mathcal{N}^2 \#(x,n,w')$. Here $W(n,w') = \{w'' w' : |w'' w'|=n\}$ is the set of length-$n$ bitstrings with suffix $w'$, $\mathcal{N} = 2^{(n'-n)/2}$ is a normalisation constant with $n' = |w'|$ and $\ket{+} = (\ket{0} + \ket{1})/\sqrt{2}$. 

We call this process a {\em quantum counter} since its outcome probabilities encode $\#(x,n,w')$ and thus $(x,n,w')$ is solved by its strong simulation. 
It is essential that the simulation is strong, as the prefactor $\mathcal{N}^2$ means that probabilities must be calculated to a precision exponential in $n$. Generally, an exponential number of weak simulations would be required to achieve this precision~\cite{Jozsa2003,VandenNest2010}.

To ensure no inefficiencies are hidden in the construction of the counters, we insist that for every $(x,n,w')$ a description of the counter can be generated in a time $\le p(|(x,n,w')|)$.
A family of counters obeying this restriction is called (polynomial-time) uniform. 
The uniformity of a family of counters is ensured if we impose two conditions on the family $\{ C_{x,n} \}$ of solution-checking circuits on which they are based: circuits in the family are of polynomial size, i.e. $N,M \le p(n)$ for each $C_{x,n}$, and the family is uniform, i.e. a description of each $C_{x,n}$ can be generated in a time $\le p(|x|,n)$.
In this article all families of counters are uniform and all families of circuits are polynomially-sized and uniform. 

Given these restrictions, a search problem is efficiently solvable if there is a family of counters that are also efficiently strongly simulable. 
An efficient strong simulation method will efficiently strongly simulate some families of counters, and thus efficiently solve some counting and search problems. This is one of the main implications of our formulation. 

In the remainder of this article we will identify counting and search problems efficiently solved by known and newly-devised strong simulation methods. However, we first describe the tensor network representation of strong simulation, which provides a convenient language for discussing these strong simulation methods.

{\bf Counting by contracting.}
The strong simulation of a quantum counter reduces to calculating the probability
\begin{equation}\label{eq:prob}
\mathcal{P} = \bra{W(n,w')} \hat{C}^{ \dagger}_{x,n} \hat{\Pi} \an{C}_{x,n} \ket{W(n,w')} .
\end{equation}
The $N$-qubit states and operators appearing on the right hand side of \eqr{eq:prob} are made up of single-qubit states and operators acting on a bounded number of qubits, according to Eqs.\ (\ref{eq:circuit})-(\ref{eq:superposition}). As is common in quantum mechanics, we may represent these states and operators by vectors and matrices, respectively (see methods). In this way, the operation $\hat{\Pi}$ and each of the $M$ operations that constitute $\hat{C}_{x,n}$ is described by a matrix of bounded size. Similarly, the initial product state $\ket{W(n,w')}$ is described by $N$ vectors of bounded size. Thus there is a set of $2(M+N)+1$ tensors that together efficiently describe the quantum process. Contracting these tensors together in the correct arrangement to obtain \eqr{eq:prob} reveals $\# (x,n,w')$. Hence we call the tensor network, illustrated in \fir{fig:fig2}, a {\em tensor counter}. 

A family of tensor counters constructed in this way inherits its uniformity from the family of quantum counters on which it is based. In this representation then, a search problem and its corresponding counting problem are efficiently solvable if there is a family of tensor counters that are efficiently contractable. This is ensured if each counter can be contracted in a time $\le p(n)$.

{\bf Identifying efficiently solvable search problems.}
We now discuss cases in which counters of the type in \fir{fig:fig2} are efficiently contractable and thereby identify efficiently solvable counting and search problems. 
Since the other parts of the counters are fixed, the efficiency of their contraction depends only on the family $\{ C_{x,n} \}$ of solution-checking circuits. 
This leads to the following result: 
a search problem is efficiently solvable if its solutions can be checked by restricted circuit families, specifically circuit families which correspond to efficiently contractable counters.

Before we identify such circuit families, let us comment on the applicability of our results to various circuit types.
We have so far outlined one approach to constructing a family of tensor counters of the form shown in \fir{fig:fig2}, based on a family $\{ C_{x,n} \}$ of quantum solution-checking circuits.
However, as we show in the methods section, there are other equally valid approaches. As an example, we describe an explicit construction for a family of tensor counters based on a family $\{ C_{x,n} \}$ of classical Boolean solution-checking circuits. 
Tensor counters can also be constructed from circuits $\{ C_{x,n} \}$ within other models of computation. For example, stochastic circuits~\cite{Johnson2010} and
atemporal Boolean circuits~\cite{Biamonte2010}, both of which are a generalisation of Boolean circuits.

Thus what follows applies to tensor counters built from solution-checking circuits $\{ C_{x,n} \}$ within any such computational model. This increases the number of routes through which one may devise a family of efficiently contractible counters, and demonstrate the efficiency of finding and counting solutions to a search problem. For notational convenience, we treat the solution-checking circuit $C_{x,n}$ and the tensor network representing it as synonymous, only making a distinction when needed for clarity. Methods for contracting tensor networks are divided into two types; geometric and algebraic.

{\em Contracting geometrically.} 
Geometric methods sequentially perform a contraction while carefully choosing the contraction sequence to avoid storing or manipulating large tensors. Accordingly, when discussing these methods, the only concern is with the structure of the counter. The counter inherits the same structure as $C_{x,n}$ up to a bounded increase in the size of the tensors, as illustrated in \fir{fig:fig3}({\bf a}). Therefore the efficiency of geometric methods depend solely on the geometries of $\{ C_{x,n} \}$.

As an example of a geometric contraction strategy, consider the case where each of the circuits in $\{ C_{x,n} \}$ has a simple tree structure. The counters can then be efficiently contracted by starting at the branch tips and moving inwards, as indicated in \fir{fig:fig3}({\bf a}). Thus search problems are efficiently solvable if their solutions can be checked by a family $\{ C_{x,n} \}$ of circuits whose gates are arranged in a simple tree structure. Markov and Shi generalised this contraction strategy by showing how to efficiently contract networks that are sufficiently tree-like~\cite{Markov2008}.
Specifically, a family of networks is efficiently contractable if each network has a treewidth (a measure of how far a network is from being a tree) growing at most logarithmically in the number of tensors. It follows from our formalism that this same contraction strategy efficiently solves any search problem whose solutions are checked by a family $\{ C_{x,n} \}$ of circuits with a treewidth growing at most logarithmically in $n$. 

We can immediately apply this result to the problem of finding satisfying solutions to a restricted version of the Boolean satisfiability decision problem~\cite{Freuder1985}. For this search problem, each instance $x$ describes a Boolean formula and each solution $w$ is a length-$n$ bitstring representing a set of $n$ Boolean variables which satisfy this formula. Provided we restrict ourselves to formulas represented by circuits $\{ C_{x,n} \}$ with a treewidth growing at most logarithmically in $n$, solutions are checked using such restricted circuits. Hence such restricted SAT-based counting and search problems are efficiently solvable using our approach.

{\em Contracting algebraically.} 
Other contraction strategies rely on algebraic relations to simplify or fully contract a tensor network; in this case it is not only the geometry of the network that is important, but also the components of the tensors. Two prominent examples are the methods of simulating stabiliser~\cite{Gottesman1999,Clark2008} and Gaussian circuits~\cite{Bravyi2005,Jozsa2008}, the latter of which encapsulates matchgate and free Fermion circuits~\cite{Valiant2001,Knill2001,Terhal2002,DiVincenzo2005}. Each is based on a group of operators which, under the action of a restricted circuit, maps to another operator in the group. Algebraic relations allow this mapping to be calculated efficiently. 

Specifically, for either a stabiliser or Gaussian circuit $C_{x,n}$, the evolved projection operator $\hat{C}^{\dagger}_{x,n}\hat{\Pi} \hat{C}_{x,n}$, appearing in the counter shown in \fir{fig:fig2}, can be found in a time $\le p(n)$~\cite{Clark2008,Jozsa2008}. This evolved projector is described by a particularly simple form of tensor network with $\le p(n)$ tensors of bounded size arranged in a linear geometry (see methods). This network, called a matrix product operator~\cite{McCulloch2007,Clark2010}, is illustrated in \fir{fig:fig3}({\bf b}). Using the strategy suggested in this figure, the simple geometry of the matrix product operator allows the full counter to be contracted in a time $\le p(n)$. Thus a search problem is efficiently solvable if its solutions can be checked by a family $\{ C_{x,n} \}$ of stabiliser or Gaussian circuits. 

For stabiliser circuits, this has a clear interpretation in terms of a known class of decision problems. A decision problem is defined as a collection of instances the solution to which is either $\mathrm{ {\tt yes}}$ or $\mathrm{ {\tt no}}$. The problem is fully specified by the set of all $\mathrm{ {\tt yes}}$ instances, called the language, and the bitstrings describing $\mathrm{ {\tt yes}}$ instances are words in this language. The class $\oplus \mathrm{L}$ contains every language for which it can be decided whether any word is a part of that language using a stabiliser circuit acting on computational basis states followed by a single measurement in that basis~\cite{Damm1990,Aaronson2004}. Using our formalism, it is then efficient to count and find words from any language in $\oplus \mathrm{L}$. 

Our formalism also allows the efficient finding of words in languages decided by a Gaussian circuit followed by a computational basis measurement in a single qubit. It turns out that such languages are trivial; whether or not a word is part of the language can be decided by considering at most a single bit of the word~\cite{VandenNest2011}.

{\em Contracting a concatenation.} Having presented several strong simulation methods in the unifying language of tensor networks, we now show that they can be combined. We have already shown that evolving $\hat{\Pi}$ by a stabiliser or Gaussian circuit $C^{[2]}_{x,n}$ returns a matrix product operator. In \fir{fig:fig3}({\bf c}) we depict how a bounded matrix product operator evolved by another circuit $C^{[1]}_{x,n}$ leads to a network with the same geometry of $C^{[1]}_{x,n}$ but with a bounded increase in the size of the tensors.  This means it is possible to contract each counter in \fir{fig:fig2} in a time $\le p(n)$ if $C_{x,n}$ is a concatenation of a circuit $C^{[1]}_{x,n}$ with a treewidth growing at most logarithmically in $n$ followed by a stabiliser or Gaussian circuit $C^{[2]}_{x,n}$.
Thus we have arrived a key result: a search problem is efficiently solvable if its solutions are checked by a family $\{ C_{x,n} \}$ of such concatenated circuits. The class of search problems of this type was not previously known to be efficiently solvable.

{\bf Power of efficiently strongly simulable circuits.} The above joint algebraic and geometric contraction strategy encapsulates each of the three strong simulation algorithms considered (stabiliser, Gaussian and tree-based) and in their combination leads to a new class of efficiently simulable quantum circuits. 

It is known that there is a sudden shift in computational power when a seemingly innocuous gate is added to the gatesets of either Gaussian or stabiliser circuits: Since they are efficiently simulable, (uniform and polynomially-sized) Gaussian or stabiliser circuits followed by a computational basis measurement cannot be used to decide, even with a bounded error, any languages outside the complexity class $\mathrm{P}$, where $\mathrm{P}$ is the class of decision problems solved efficiently and without error using a classical computer. However, if SWAP~\cite{Jozsa2008} or the $\pi/8$-gate~\cite{Nielsen2000}, respectively, are added to either gateset, then either can decide problems in $\mathrm{BQP}$. This is the class of decision problems solved efficiently and with a bounded error using a quantum computer.

The shift is enhanced by both our concatenation of simulation techniques and our algorithm for searching by simulating. The former narrows down where the shift may take place: allowing bounded-size, but otherwise arbitrary gates, in a limited geometry before Gaussian or stabiliser circuits does allow the deciding of languages beyond $\mathrm{P}$ with bounded error. The latter highlights the size of the shift: we expect the languages that can be decided without error using Gaussian, stabiliser, tree-like circuits, or their concatenation followed by a computational basis measurement not only to be a subset of $\mathrm{P}$, but a proper subset. 
To see why this is expected, consider the opposite to be true. 
This would imply that all efficiently-checkable search problems can be checked using Gaussian, stabiliser, tree-like circuits, or their concatenation followed by a computational basis measurement. Using our formulation, it would then be possible to efficiently count and find solutions to all efficiently-checkable search problems. Note that each instance of a problem in $\# \mathrm{P}$ ($\mathrm{NP}$) corresponds to counting the number of solutions (deciding whether there is a solution) to an instance of an efficiently-checkable search problem~\cite{Goldreich2010}. Thus the complexity hierarchy $\mathrm{P} \subseteq \mathrm{NP} \subseteq \# \mathrm{P}$ would collapse to an equality.

\section*{Discussion}
Our results reveal a hitherto unknown consequence of algorithms that efficiently simulate quantum circuits; each allows for the efficient solving of a class of search problems. 
This applies to every strong simulation technique, implemented by any device, classical or quantum. 
For well understood methods such as stabiliser circuit simulation algorithms, this confirms what is already known. For methods whose computational power is less well understood, this demonstrates the efficient solubility of several classes of search problems based on how their solutions are checked. To our knowledge, this extends the class of counting and search problems known to be efficiently solvable.
Further, the most general technique considered in this article corresponds to a novel combination of strong simulation methods. Hence this article widens the class of efficient strong simulation methods and adds to their applicability.

\section*{Methods}
{\bf Generic quantum counters.}
Here we give an explicit approach to constructing a quantum solution-checking circuit $C_{x,n}$ used to form a quantum counter.

For every instance $x$ of an efficiently-checkable search problem a solution of length $n$ can be checked in time $\le p(|x|)$ using a classical computer. Thus there is a classical algorithm that decides the language $\mathcal{L} = \{(x, w) : w \in S(x) \}$ in time $\le p(|(x,w)|)$.
It is then possible to construct a family of Boolean circuits $\{ C_m \}$ with the following properties~\cite{Arora2009}: each $C_m$ takes inputs $(x,w)$ with $|(x,w)| = m$ and outputs logical-one if $(x,w) \in \mathcal{L} $, and otherwise logical-zero; it comprises $\le p(m)$ AND, OR and NOT gates~\cite{Nielsen2000}; and it is constructible in a time $\le p(m)$. Hardwiring $x$ into $C_m$ leaves a Boolean solution-checking circuit $C_{x,n}$ that takes bitstrings $w$ of length $n$ and outputs logical-one if $w \in S(x)$, and otherwise logical-zero.

The Boolean circuit $C_{x,n}$ can be made reversible in time $\le p(|x|,n)$ with only $\le p(|x|,n)$ additional gates and $\le p(|x|,n)$ additional ancilla bits, each initially set to zero~\cite{Saeedi2011,Bennett1973,Toffoli1980}. 
The standard mapping between reversible classical and unitary quantum circuits~\cite{Nielsen2000} leads to a quantum circuit of the desired form. This circuit is generated in time $\le p(|x|,n)$ and comprises $\le p(|x|,n)$ gates. So the approach leads to polynomially-sized uniform families of quantum solution-checking circuits, and hence a uniform family of quantum counters for the search problem.

{\bf Tensor counters from quantum counters.}
Here we describe how the expression for the probability $\mathcal{P}$ in \eqr{eq:prob} is replaced by the contraction of a uniform tensor network. Other descriptions of how to represent quantum circuits by tensor networks can be found in Refs.~\cite{Jozsa2006,Markov2008}.

To begin, insert Eqs.\ (\ref{eq:circuit})-(\ref{eq:superposition}) into \eqr{eq:prob} to obtain
\begin{widetext}
\begin{equation}\label{eq:prob2}
\mathcal{P} = 
\bra{0}^{\otimes (N-n)} \bra{ + }^{\otimes (n-n')} \bra{ w'_{n'} } \cdots \bra{ w'_1 } 
\left( \prod_{k=M}^1 \cre{g}_k \right)
\ket{1} \bra{1}
\left( \prod_{k=1}^M \an{g}_k \right)
\ket{0}^{\otimes (N-n)} \ket{ + }^{\otimes (n-n')} \ket{ w'_{n'} } \cdots \ket{ w'_1 } .
\end{equation}
\end{widetext}

To rewrite this as a tensor network contraction, first replace the states and operators $\ket{0}$, $\ket{1}$, $\ket{+}$, $\{ \hat{g}_k \}$ and $ \ket{1} \bra{1}$ appearing in \eqr{eq:prob2} by their corresponding tensors $[0]$, $[1]$, $[+]$, $\{ [g_k] \}$, and $[\ket{1} \bra{1}]$. Specifically, represent each single-qubit state $\ket{ \psi}$ by the tensor $[\psi]$ with $2$ components $[\psi]^{ i_1} = \braket{i_1}{\psi}$ labelled by an index $i_1 = \{0 ,1 \}$. To represent the dual of a state, take the conjugate transpose. Further, represent each $N'$-qubit operator $\hat{O}$ by the tensor $[O]$ with $2^{2N'}$ components 
\begin{align}
[O]^{o_{N'} \dots o_1 i_{N'} \dots i_1} = \bra{i_{N'}} \cdots \bra{i_1} \hat{O} \ket{o_{N'}} \cdots \ket{o_1} , \nonumber
\end{align} 
labelled by $2N'$ indices, where $o_j = \{0 ,1 \}$. After this replacement, write a joint summation (contraction) over every pair of indices corresponding to an input and output connected by the expression in the right hand side of \eqr{eq:prob2}. 

It follows from above that the mapping can be performed in a time $\le p(|(x,n,w')|)$. Thus a uniform family of quantum counters maps to a uniform family of tensor counters.

{\bf Beyond quantum circuits.}
Here we describe conditions on the tensor network representing $C_{x,n}$, which appears in the counter shown in \fir{fig:fig2}. The only requirement needed to ensure the contraction of the counter returns the desired value of $\mathcal{P}$ is
\begin{align} \label{eq:condition}
\sum_{o_N,\dots,o_2} | [C_{x,n}]^{0 \dots 0 w  o_N \dots o_2 1} |^2
 =  \left\{ 
  \begin{array}{l l}
    1 & \text{$w \in S(x)$}\\
    0 & \text{otherwise}\\
  \end{array} \right. 
,
\end{align}
where $[C_{x,n}]$ is the tensor representing $\hat{C}_{x,n}$. In fact, it is not necessary for $[C_{x,n}]$ to have the same number $N$ of output indices as input indices. This gives us the freedom to consider a tensor network that does not represent a quantum circuit, but nevertheless whose contraction returns a tensor $[C_{x,n}]$ satisfying \eqr{eq:condition}. 

{\bf Counters based on Boolean circuits.}
Here we give an explicit approach to constructing a tensor counter built from a classical Boolean circuit. 

In our construction of a quantum counter, we showed how to construct a solution-checking Boolean circuit $C_{x,n}$ for any solution-length $n$ and instance $x$ of a given search problem. The circuit can be generated in a time $\le p(|x|,n)$ and comprises $\le p(|x|,n)$ bounded-size gates. 
It implements the binary switching function $f_{x,n} : \{0,1\}^{n} \to \{0,1\}$ where $f_{x,n} (w)$ is 1 if $w \in S(x)$, and otherwise 0.

To obtain a tensor network representing $C_{x,n}$, first represent each gate implementing the binary function $g_k : \{0,1\}^{n_k} \to \{0,1\}^{m_k}$ by a tensor with components $[g_k]^{i_k o_k} = \delta_{g_k (i_k),o_k}$ where $i_k = i_{n_k} \cdots i_1$ and similarly for $o_k$. Next connect the input and output indices of the tensors $\{ [g_k] \}$ in precisely the same arrangement as the circuit. The components of the tensor obtained by contracting this network are $[C_{x,n}]^{i o} = \delta_{f_{x,n} (i),o}$. Again we have used the shorthand $i = i_{n} \cdots i_1$. This tensor satisfies \eqr{eq:condition}. Thus the network representing $C_{x,n}$ leads to a valid tensor counter of the form shown in \fir{fig:fig2}. The counter is generated in a time $\le p(|(x,n,w')|)$ and so counters constructed in this way form a uniform family.

{\bf Contracting algebraically.}
Using the Heisenberg picture, stabiliser circuits map a product of Pauli operators to another Pauli product~\cite{Clark2008}. Such a product is represented by a particularly simple matrix product operator where each tensor is the standard matrix representing the corresponding Pauli operator and the internal indices have dimension $\chi = 1$. With $m$ the number of qubits, the projector $\hat{\Pi} = \half (\mathbbm{1}_2 - \sigma_z) \mathbbm{1}_2^{\otimes (m-1)} $ is the sum of two Pauli products, and therefore so is $\hat{C}^{\dagger}_{x,n}\hat{\Pi} \hat{C}_{x,n}$. Two matrix product states can be efficiently summed to make another where the dimension $\chi$ increases additively~\cite{McCulloch2007,Clark2010}. Thus $\hat{C}^{\dagger}_{x,n}\hat{\Pi} \hat{C}_{x,n}$ is represented by a $\chi = 2$ matrix product operator.

Gaussian circuits map generators $c_\mu$ of a Clifford algebra to a linear combination $\sum_\nu \tilde{R}_{\mu \nu} c_\nu$ where $\tilde{R}_{\mu \nu}$ is calculable in a time $\le p(m)$ and $\nu$ takes $2m$ values~\cite{Jozsa2008}. Each $c_\mu$ may be represented by a Pauli product~\cite{Jozsa2008}, and so $\hat{C}^{\dagger}_{x,n} c_\mu \hat{C}_{x,n}$ by a $\chi = 2m$ matrix product operator. A product of two generators $c_{1} c_{2}$ is mapped to $\hat{C}^{\dagger}_{x,n} c_{1} \hat{C}_{x,n} \hat{C}^{\dagger}_{x,n} c_{2} \hat{C}_{x,n}$, represented by the product of two matrix product operators. The result is another matrix product operator where the dimension $\chi$ increases multiplicatively~\cite{McCulloch2007,Clark2010}. In this case the result has dimension $\chi = (2m)^2$. Finally, there is a choice of $c_{1}$ and $c_{2}$ such that $\hat{\Pi} = \half (\mathbbm{1}_{2^m} - c_{1} c_{2})$~\cite{Jozsa2008}. Thus $\hat{C}^{\dagger}_{x,n}\hat{\Pi} \hat{C}_{x,n}$ is represented by a $\chi = (2m)^2 + 1$ matrix product operator.

\section*{Acknowledgements}
THJ thanks Vlatko Vedral for helpful discussions, the Centre for Quantum Technologies in Singapore, where much of this work was completed, and Federica Ferraris for help drawing the figures.

\begin{figure*}[p]
\includegraphics[width=18cm]{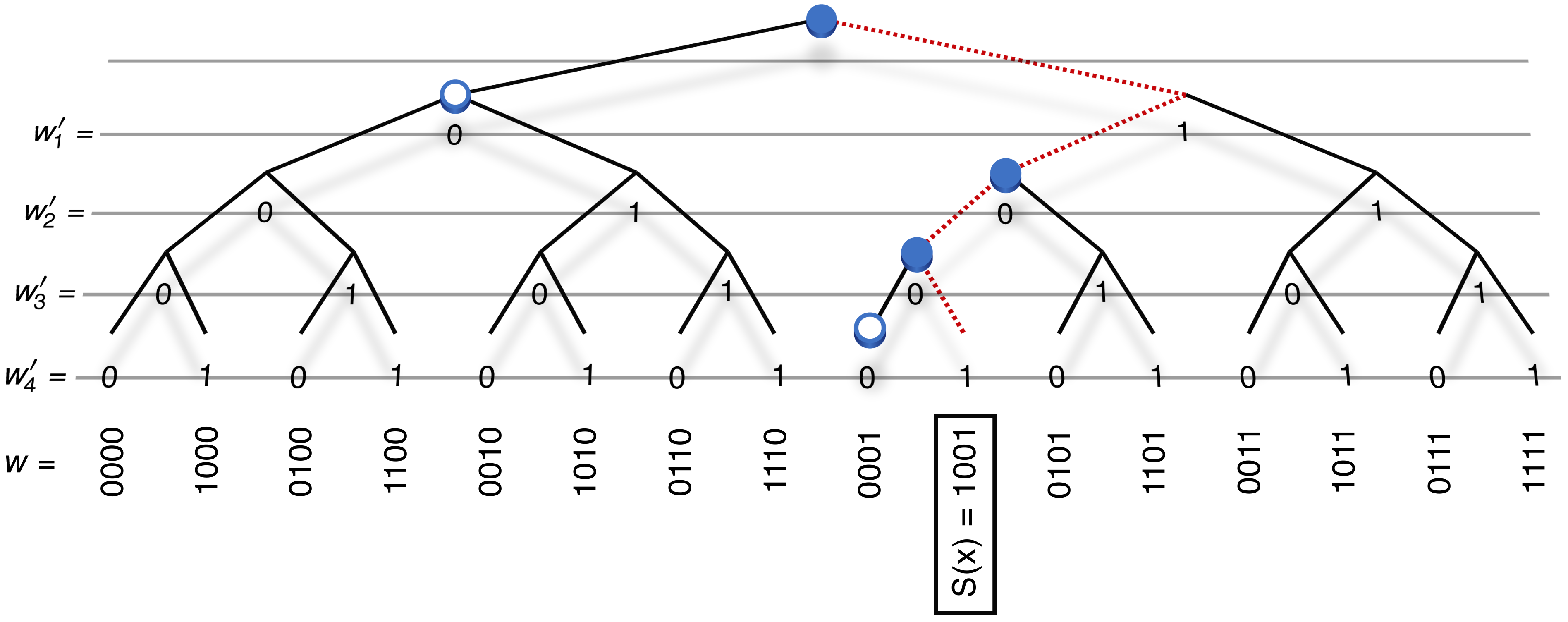}
\caption{\label{fig:fig1} {\bf Searching by counting}.
A solution to a search problem instance $x$ can be inferred from a sequence of a small number of counts. In the example shown there is a single solution $S(x) = 1001$ and each blue circle represents an enumeration of solutions to $x$ of length $4$ ending in suffix $w'$. An empty or filled circle indicates that the count returned zero or one, respectively. The red dotted line shows the path to the solution inferred from these counts. For the first count (upper most circle), $w'$ has zero length, i.e., all solutions of length 4 are counted and we infer there is a solution of this length. For the next count $w' = w'_1 = 0$ and we infer that solutions must begin with suffix $1$. For the third count $w' = w'_2 w'_1  = 01$ and we infer that there is a solution beginning with suffix $01$, and so on.}
\end{figure*}

\begin{figure}[p]
\includegraphics[width=8.8cm]{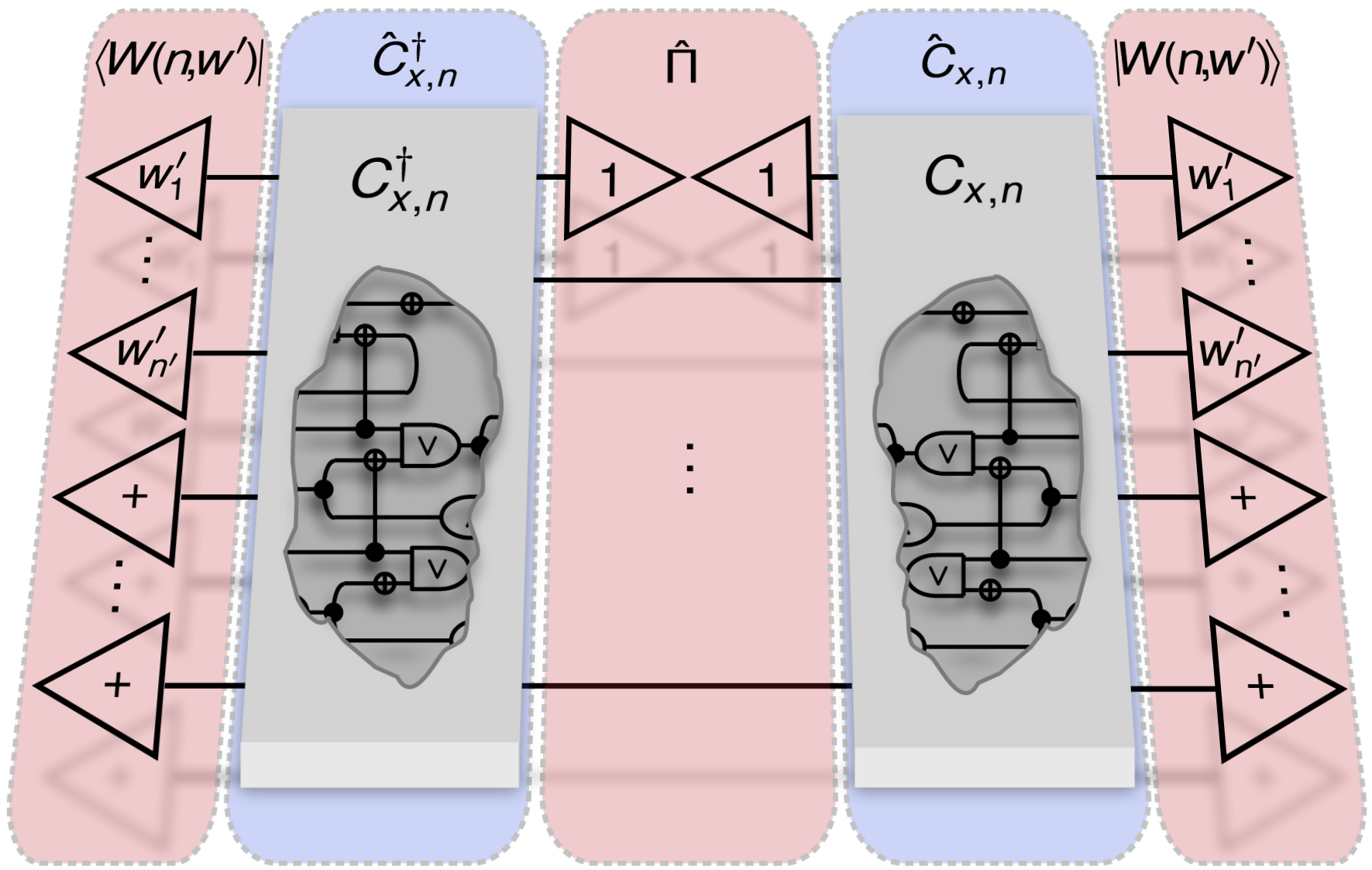}
\caption{\label{fig:fig2} {\bf Tensor counter}.
A tensor network whose contraction reveals the solution to a counting problem instance $(x,n,w')$. 
Each shape represents a tensor, and each line leaving it corresponds to one of its indices. A connection between two shapes represents a contraction, or joint sum over the corresponding indices.
This tensor counter generalises the expression in \eqr{eq:prob}, written at the top of the figure, for the probability of a measurement outcome following a quantum circuit. 
For clarity, tensors representing ancilla bits have been absorbed into the network representing the solution-checking circuit $C_{x,n}$. This circuit may be quantum, stochastic or Boolean (shown above).}
\end{figure}

\begin{figure*}[p]
\includegraphics[width=18cm]{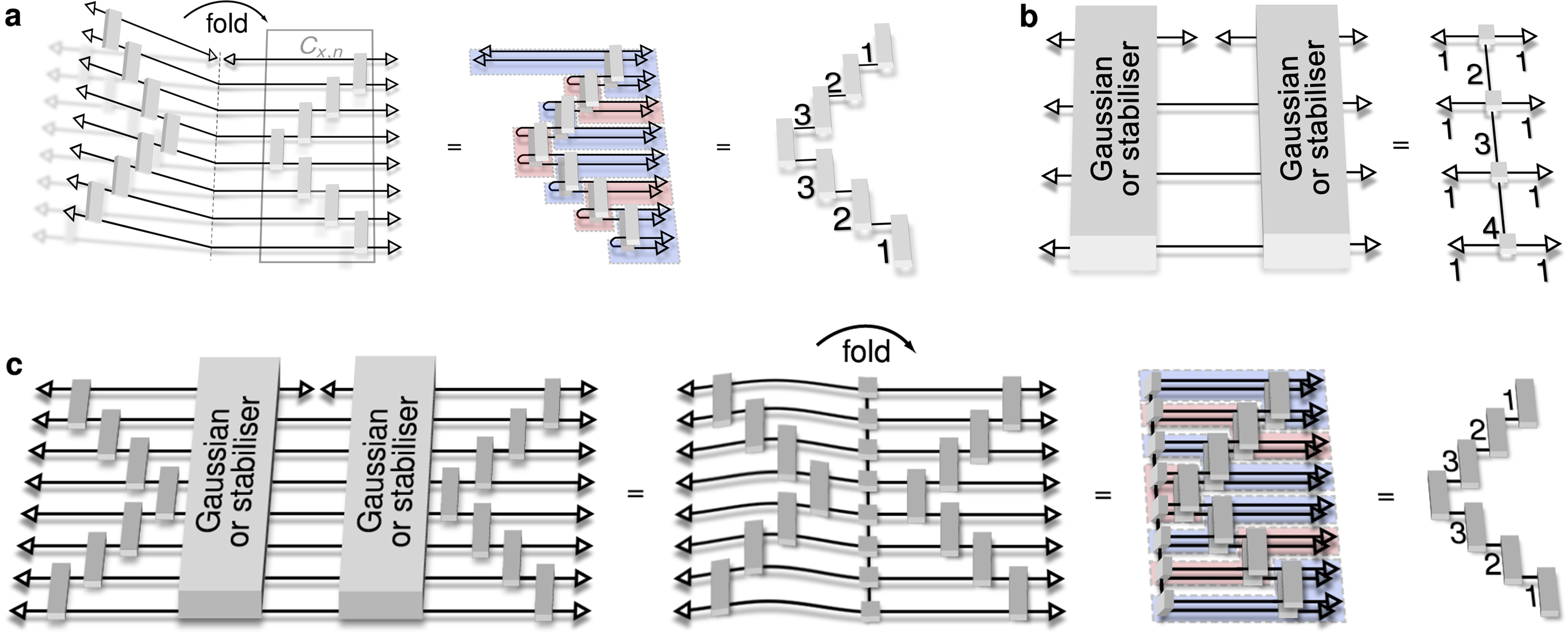}
\caption{\label{fig:fig3} {\bf Contraction strategies}.
({\bf a}) Folding a tensor counter and merging the tensors in each of the blue or red shaded regions reduces the counter to a network with the geometry of the solution-checking network $C_{x,n}$. If $C_{x,n}$ has a tree structure, as is shown, then the counter can be contracted efficiently in the order labelled, starting from $1$. ({\bf b}) A Gaussian or stabiliser circuit acting on the local projection operator returns a matrix product operator of bounded size. This can be efficiently contracted with the initial state in the order labelled. ({\bf c}) If $C_{x,n}$ is the concatenation of a tree-like circuit followed by a Gaussian or stabiliser circuit then the counter may be efficiently contracted in the following way. Firstly, as in ({\bf b}), the action of the Gaussian or stabiliser circuit returns a matrix product operator. Secondly, as in ({\bf a}), folding and merging
returns a network with the same geometry of the tree-like circuit, which may be efficiently contracted.}
\end{figure*}


\begin{thebibliography}{99}

\bibitem{Shor1994}
Shor, P. Algorithms for quantum computation: Discrete logarithms and factoring, {\em SIAM J. Comput.} {\bf 26,} 1484--1509 (1997).

\bibitem{VandenNest2010}
Van den Nest, M. Classical simulation of quantum computation, the Gottesman-Knill theorem, and slightly beyond. {\em Quant. Inf. Comp.} {\bf 10,} 0258--0271 (2010).

\bibitem{Jozsa2003}
Jozsa, R. \& Linden, N. On the role of entanglement in quantum-computational speed-up. {\em Proc. R. Soc. Lond. A} {\bf 459,} 2011--2032 (2003).

\bibitem{Gottesman1999}
Gottesman, D. A theory of fault-tolerant quantum computation. {\em Phys. Rev. A} {\bf 57,} 127137 (1998).

\bibitem{Aaronson2004}
Aaronson, S. \& Gottesman, D. Improved simulation of stabilizer circuits. {\em Phys. Rev. A} {\bf 70,} 052328 (2004).

\bibitem{Anders2006}
Anders, S. \& Briegel, H. J. Fast simulation of stabilizer circuits using a graph-state representation. {\em Phys. Rev. A} {\bf 73,} 022334 (2006).

\bibitem{Clark2008}
Clark, S., Jozsa, R. \& Linden, N. Generalized Clifford groups and simulation of associated quantum circuits. {\em Quant. Inf. Comp.} {\bf 8,} 0106--0126 (2008).

\bibitem{Valiant2001}
Valiant, L. G. Quantum computers that can be simulated classically in polynomial time. {\em SIAM J. Comput.} {\bf 31,} 1229--1254 (2002).

\bibitem{Knill2001}
Knill, E. Fermionic linear optics and matchgates. Preprint at $<$http://arxiv.org/abs/quant-ph/0108033$>$ (2001).

\bibitem{Terhal2002}
Terhal, B. M. \& DiVincenzo, D. P. Classical simulation of noninteracting-fermion quantum circuits. {\em Phys. Rev. A} {\bf 65,} 032325 (2002).

\bibitem{DiVincenzo2005}
DiVincenzo, D. P. \& Terhal, B. M. Fermionic linear optics revisited. {\em Found. Phys.} {\bf 35,} 1967--1984 (2005).

\bibitem{Bravyi2005}
Bravyi, S. Lagrangian representation for fermionic linear optics. {\em Quantum Inf. Comp.} {\bf 5,} 216--238 (2005).

\bibitem{Jozsa2008}
Jozsa, R. \& Miyake, A. Matchgates and classical simulation of quantum circuits. {\em Proc. R. Soc. A} {\bf 464,} 3089--3106 (2008).

\bibitem{Vidal2003}
Vidal, G. Efficient classical simulation of slightly entangled quantum computations. {\em Phys. Rev. Lett.} {\bf 91,} 147902 (2003).

\bibitem{Jozsa2006}
Jozsa, R. On the simulation of quantum circuits. Preprint at $<$http://arxiv.org/abs/quant-ph/0603163$>$ (2006).

\bibitem{Shi2006}
Shi, Y.-Y., Duan, L.-M. \& Vidal, G. Classical simulation of quantum many-body systems with a tree tensor network. {\em Phys. Rev. A} {\bf 74,} 022320 (2006).

\bibitem{Yoran2006}
Yoran, N. \& Short, A. J. Classical simulation of limited-width cluster-state quantum computation. {\em Phys. Rev. Lett.} {\bf 96,} 170503 (2006).

\bibitem{Yoran2007}
Yoran, N. \& Short, A. J. Efficient classical simulation of the approximate quantum Fourier transform. {\em Phys. Rev. A} {\bf 76,} 042321 (2007).

\bibitem{Browne2007}
Browne, D. E. Efficient classical simulation of the quantum Fourier transform. {\em New J. Phys.} {\bf 9,} 146 (2007).

\bibitem{Bravyi2007}
Bravyi, S. \& Raussendorf, R. On measurement-based quantum computation with the toric code states. {\em Phys. Rev. A} {\bf 76,} 022304 (2007).

\bibitem{VandenNest2007}
Van den Nest, M., D\"{u}r, W., Vidal, G. \& Briegel, H. J. Classical simulation versus universality in measurement-based quantum computation. {\em Phys. Rev. A} {\bf 75,} 012337 (2007). 

\bibitem{Markov2008}
Markov, I. L. \& Shi, Y.-Y. Simulating quantum computation by contracting tensor networks. {\em SIAM J. Comput.} {\bf 38,} 963--981 (2008).

\bibitem{Bravyi2009}
Bravyi, S. Contraction of matchgate tensor networks on non-planar graphs. {\em Contemporary Mathematics} {\bf 482,} 179--211 (2009).

\bibitem{VandenNest2009}
Van den Nest, M., D\"{u}r, W., Raussendorf, R. \& Briegel, H. J. Quantum algorithms for spin models and simulable gate sets for quantum computation. {\em Phys. Rev. A} {\bf 80,} 052334 (2009).

\bibitem{Biamonte2010}
Biamonte, J. D., Clark, S. R. \& Jaksch, D. Categorical tensor network states. {\em AIP Advances} {\bf 1,} 042172 (2011).

\bibitem{Johnson2010}
Johnson, T. H., Clark, S. R., \& Jaksch, D. Dynamical simulations of classical stochastic systems using matrix product states. {\em Phys. Rev. E} {\bf 82,} 036702 (2010).

\bibitem{Damm1990}
Damm, C. Problems complete for $\oplus$L. {\em Information Processing Letters} {\bf 36,} 247--250 (1990).

\bibitem{Cai2007}
Cai, J.-Y. \& Lu, P. Holographic algorithms: From art to science. {\em J. Comput. Syst. Sci.} {\bf 77,} 41--61(2011).

\bibitem{Goldreich2010}
Goldreich, O. {\em P, NP, and NP-completeness: The basics of computational complexity}, (Cambridge Univ. Press, 2010).

\bibitem{Freuder1985}
Freuder, E. C. A sufficient condition for backtrack-bounded search. {\em J. ACM} {\bf 32,} 755--761 (1985).

\bibitem{McCulloch2007}
McCulloch, I. From density-matrix renormalization group to matrix product states. {\em J. Stat. Mech.} P10014 (2007).

\bibitem{Clark2010}
Clark, S. R., Prior, J., Hartmann, M. J., Jaksch, D. \& Plenio, M. B., Exact matrix product solutions in the Heisenberg picture of an open quantum spin chain. {\em New J. Phys.} {\bf 12,} 025005 (2010).

\bibitem{VandenNest2011}
Van den Nest, M. Quantum matchgate computations and linear threshold gates. {\em Proc. R. Soc. A} {\bf 467,} 821--840 (2011).

\bibitem{Nielsen2000}
Nielsen, M. \& Chuang, I. L. {\em Quantum computation and quantum information}, (Cambridge Univ. Press, 2000).

\bibitem{Arora2009}
Arora, S. \& Barak, B. {\em Computational complexity: A modern approach}, (Cambridge Univ. Press, 2009).

\bibitem{Bennett1973}
Bennett, C. H. Logical reversibility of computation. {\em IBM J. Res. Develop.} {\bf 17,} 525--532 (1973).

\bibitem{Toffoli1980}
Toffoli, T. Reversible computing. {\em Automata, languages and programming, Seventh colloquium, Lecture notes in computer science} {\bf 84,} de Bakker, J. W. \& van Leeuwen, J. eds., 632Ð644 (Springer, 1980).

\bibitem{Saeedi2011}
Saeedi, M. \& Markov, I. L. Synthesis and optimization of reversible circuits - A Survey. $<$http://arxiv.org/abs/1110.2574$>$ (2011).

\end{thebibliography}
\end{document}